\documentstyle[12pt]{article}
\newcommand{\bfl}{\begin{flushleft}}
\newcommand{\efl}{\end{flushleft}}
\newcommand{\bc}{\begin{center}}
\newcommand{\ec}{\end{center}}
\newcommand\ie {{\it i.e. }}

\newcommand\jump{\vspace*{17pt}}

\def\be{\begin{eqnarray}}
\def\ee{\end{eqnarray}}
\newenvironment{draftequation}[1]{\be\label{#1}}{\ee} 
\newcommand\bbe[1]{\begin{draftequation}{#1}}
\newcommand\eee{\end{draftequation}}

\def\half{{\textstyle{1 \over 2}}}
\def\ihalf{{\textstyle{i \over 2}}}

\begin{document}

\begin{flushright}
USITP-93-12\\
May  1993\\
hep-th/9307108
\end{flushright}
\bigskip
\Large
\begin{center}
\bf{Classical and Quantized Tensionless Strings}\\

\bigskip

\normalsize
by\\
\bigskip

J. Isberg\footnote{udah267@elm.cc.kcl.ac.uk},\\
{\it Department of Mathematics,\\
King's College London,\\
Strand, London WC2R 2LS\\
ENGLAND\\
\bigskip
and\\}
\bigskip
U. Lindstr\"om\footnote{ul@vana.physto.se},
B. Sundborg\footnote{bo@vana.physto.se} and G. Theodoridis\\
{\it Institute of Theoretical Physics\\
University of Stockholm\\
Box 6730\\
S-113 85 Stockholm\\
SWEDEN}\\
\end{center}
\vspace{1.0cm}
\normalsize
{\bf Abstract:} From the ordinary tensile string we derive a geometric
action for the tensionless ($T=0$) string and discuss its symmetries
and field equations. The Weyl symmetry of the usual string is shown to
be replaced by a global space-time conformal symmetry in the $T\to 0$
limit. We present the explicit expressions for the generators of this
group in the light-cone gauge. Using these, we quantize the theory in
an operator form and require the conformal symmetry to remain a
symmetry of the quantum theory.  Modulo details concerning zero-modes
that are discussed in the paper, this leads to the stringent
restriction that the physical states should be singlets under
space-time diffeomorphisms, hinting at a topological theory. We
present the details of the calculation that leads to this conclusion.

\bigskip

\bigskip

\begin{flushleft}
\section{Introduction}
\end{flushleft}

\bigskip

The high energy limit of string theory is still quite poorly
understood, despite many important and interesting results on high
energy scattering, \cite{AMAT3}-\cite{GROS4} and high temperature
behaviour \cite{ALVA1}-\cite{ATIC1}.
Just like the massless limit in particle theory sheds light on short
distance field theory, the zero tension limit, $T\to 0$, of strings is
expected to illuminate some short-distance properties of string
theory. In particular we hope that the intriguing high energy
symmetries discussed by Gross \cite{GROS1} may be studied in this
limit. Thereby one would presumably be able to probe the conjectured
unbroken "topological" phase of general covariance
\cite{WITT1,WITT2}.
Though much work has been done in topological field theory, \cite
{WITT3,LABA1,WITT4,BIRM1}, much less is known about the relation to
string theory. The present work supports and substantiates such a
connection.  In fact, our results point in this direction in quite an
unexpected way. We have approached the problem by first formulating
the exactly tensionless case, where additional symmetries relating
different mass levels, in particular space-time conformal symmetry,
should appear. In a previous series of papers we have studied various
aspects of the tensionless case of the bosonic string \cite{akul}, the
superstring \cite{ulbsgt1} and the spinning string
\cite{ulbsgt2,roli}. We have also reported on the main result of this paper in
a brief letter \cite{jiulbs}.

In addition to the questions concerning the deep symmetries of string
theory that may be revealed in the tensionless limit, there are
fundamental problems of string perturbation theory that can be
illuminated once a tensionless theory has become available. It was
found in \cite{mass-shift} that the string tension is unchanged by
one-loop corrections (for type II superstrings), but {\it mass level
shifts are larger than the level separation} for high levels, no
matter how small the coupling constant is. Such a behaviour is bound
to cause severe problems for the tensile string perturbation theory,
and it is probably one of the simplest manifestations of the rapid
large order growth of the string perturbation series for massless
external states discussed by Gross and Periwal
\cite{Gross-Per}. In similar problems appearing in quantum mechanics, one
turns to quasi-degenerate perturbation theory. Taken over to string
physics, this recipe would mean that we start from the degenerate case
with vanishing zeroth order level separation. But this is precisely
the limit of vanishing tension. Thus a future interacting tensionless
theory could become a key to non-perturbative string physics.

Other authors have also discussed tensionless strings and their
quantization \cite{zh,banerual,liraspsr,gararual} ever since they were
first discussed by Schild \cite{sc}.

One expects either a continuous or a massless spectrum when the scale
given by the string tension is removed from the theory. Correspondence
with classical tensionless strings would favour a continuous spectrum,
but on the other hand all $T\not= 0$ string states approach zero mass
as $T\to 0$. The result of the present work indicates that the
massless spectrum is the correct answer, but we also find extremely
restrictive constraints on the spectrum, {\it effectively allowing
only states invariant with respect to general coordinate
transformations}. One could envisage a spectrum of string states
characterized by topology, but we have not yet found a concrete
construction of a satisfactory Hilbert space.

In this paper we will be concerned with the quantization of the
tensionless closed bosonic string. In particular we will explore the
(space-time!) conformal symmetry of this model and investigate under
what condition this symmetry survives at the quantum level. When the
two-dimensional reparametrization invariance of ordinary strings is
gauge-fixed in the light-cone gauge, anomalies in the local symmetry
are reflected in a breakdown of the Lorentz algebra (in non-critical
dimensions). Similarly, inconsistencies in the quantum geometry of the
tensionless string can be probed by checking the conformal algebra in
the light-cone gauge. We are further motivated to demand space-time
conformal invariance in the quantized theory because we find that this
is the symmetry that replaces Weyl-invariance in the $T \to 0$ limit.
Hence when we find obstructions to conformal invariance, rather than
conclude that the symmetry is broken, we interpret the obstructions as
conditions on the physical states of fundamental strings. In a future
tensionless limit of $QCD$ strings, it might instead be more
appropriate to accept breakdown of conformal invariance.

We have chosen to make the article relatively self-contained and to
include some new result on the classical theory. The article is
organized as follows: In Section $2$ we present the classical theory.
We derive actions for a tensionless string (or tensionless
$p$-branes), discuss the symmetries of the string action we choose to
work with, present the equations of motion, introduce the light-cone
gauge and finally give the compensating reparametrizations needed to
stay in that gauge after applying a conformal transformation. Section
$3$ contains the quantum theory. We begin with a brief discussion of
BRST-quantization. The main part of the paper is then the detailed
derivation of the anomalies in the light-cone operator algebra and a
discussion of their implications. Finally we end the article with our
conclusions. A discussion of the relation between vacuum in the
tensile and in the tensionless model is given as an appendix.

\begin{flushleft}
\section{The classical theory}
\end{flushleft}

\bigskip
\begin{flushleft}
{\bf 2.1 Actions}
\end{flushleft}
\bigskip

In this section we discuss the classical theory of strings in the
limit that the tension $T\to 0$. At essentially no extra cost we can
discuss such a limit also for higher dimensional objects commonly
known as $p$-branes. Although the rest of the article deals
exclusively with strings, in this section we keep the discussion
general, and the formulae for strings used in other sections are
obtained by setting $p=1$.

Consider a theory given by an action of the form
\bbe{act1} S=T\int{d^{p+1}\xi {\cal L}},
\eee
e.g., a $p$-brane with space time coordinates $X^m$, world volume
coordinates $\xi ^\alpha$ and "tension" $T$. There are numerous ways
of rewriting the action so that the $T\to 0$ limit may be taken. We
will settle on a formulation that has a geometric interpretation, but
first display the simplest generalization of the point-particle
action, ($p=0$, $T=m$), involving an auxiliary field $\phi$:
\bbe{phiact}
S=\half\int{ d^{p+1}\xi
\left[{
\phi{\cal
L}^2+\phi^{-1}T^2 }\right]}.
\eee
(The equivalence is seen by integrating out $\phi$.) Here the limit
$T\to 0$ can be readily taken. With $\phi \to e$ this procedure yields
the reparametrization invariant action involving the einbein $e$ for
the massless point-particle. With $p=1$ and $T$ the string tension
this procedure was used in, e.g.,
\cite{akul} to obtain an action for the tensionless
string. It is only in the point-particle case that there is a
connection to the world-volume geometry, however. We have found it
useful to try to maintain such a relation and have therefore choosen a
different route to the $T\to 0$ limit.

The starting point is the Nambu-Goto-Dirac world volume action
\bbe{pact}
S=T\int{ d^{p+1}\xi \sqrt{-det\gamma _{\alpha\beta}} }
\eee
where $X^m=X^m (\xi)$ and
\bbe{indm}
\gamma _{\alpha\beta}\equiv \partial _\alpha X^m \partial _\beta X^n \eta
_{mn}
\eee
is the metric induced on the world volume from the Minkowski
space-time metric $\eta_{mn}$. We reformulate the theory in phase
space. The generalized momenta derived from the Lagrangian in
(\ref{pact}) are
\be
P_m =T\sqrt{-\gamma}\gamma^{\alpha 0}\partial _\alpha X_m .
\ee
where $\gamma ^{\alpha \beta}$ is the inverse of $\gamma _{\alpha
\beta}$.  They satisfy the constraints
\bbe{cons}
P^2+T^2\gamma \gamma ^{00}=0\cr P_m \partial_aX^m=0,\quad a=1,...,p.
\eee
Here $\gamma \equiv det\gamma_{\alpha\beta}$.  As usual for a
diffeomorphism invariant theory, the naive Hamiltonian vanishes and
the total Hamiltonian consists of the sum of the constraints
(\ref{cons}) multiplied by Lagrange multipliers, which we shall call
$\lambda$ and $\rho ^a$:
\bbe{hami}
{\cal H}=\lambda (P^2+T^2\gamma
\gamma^{00})+\rho^aP\cdot\partial_aX
\eee
The phase space action thus becomes
\bbe{phact}
S^{PS}=\int{d^{p+1}\xi \left\{{P\cdot \dot{X}-\lambda (P^2+T^2\gamma
\gamma^{00})-\rho^aP\cdot\partial_aX}\right\}}.
\eee
We integrate out the momenta to find the configuration space action
\bbe{cact}
S^{CS}=\half\int{d^{p+1}\xi {1 \over
{2\lambda}}\left\{{\dot{X}^2-2\rho ^a \dot X^m\partial _aX_m +\rho
^a\rho^b\partial _bX^m \partial _aX_m -4\lambda
^2T^2\gamma\gamma^{00}}\right\}}.
\eee
For $p=1$ we may identify
\bbe{poneg}
g^{\alpha \beta}=
\left( \matrix{-1\quad\hfill \rho  \cr
 \rho \quad\hfill-\rho ^2+4\lambda^2T^2\cr}
\right)
\eee
which leads to the usual Weyl invariant tensile string action
\bbe{ggact}
S=-\half T\int{d^{2}\xi\sqrt{-g}\left\{{g^{\alpha\beta}\partial
_\alpha X^m
\partial _\beta X^n \eta
_{mn}}\right\}}.
\eee
For $p>1$ it is not possible to directly identify the geometric fields
in (\ref{cact}). We first have to rewrite it as \cite{HASS}
\bbe{Gact}
S^{CS}=\half\int{d^{p+1}\xi \left\{{
{{h^{\alpha\beta}\gamma_{\alpha\beta}}\over {2\lambda}} -2\lambda
T^2G(p-1)+2\lambda T^2GG^{ab}\gamma_{ab} }\right\}}
\eee
where
\be
h^{\alpha\beta}=\left( \matrix{1\quad\hfill -\rho ^a\cr -\rho
^a\quad\hfill\rho ^a\rho ^b\cr} \right)
\ee
is a rank 1 auxiliary matrix and $G_{ij}$ is a $p$-dimensional
auxiliary metric with determinant $G$.  (Integrating out $G_{ij}$ we
recover (\ref{cact}).) Now the identification
\bbe{gdef}
g^{\alpha\beta}= {1 \over 4}T^{-2}\lambda^{-2}G^{-1}\left(
\matrix{-1\quad\hfill \rho ^a\cr \rho ^a\quad\hfill-\rho ^a\rho
^b+4\lambda^2T^2GG^{ab}\cr}
\right)
\eee
produces the usual $p$-brane action involving the world volume metric
$g_{\alpha\beta}$:
\bbe{gact}
S=-\half T\int{d^{p+1}\xi\sqrt{-g}\left\{{g^{\alpha\beta}\partial
_\alpha X^m
\partial _\beta X^n \eta
_{mn}-(p-1)}\right\}}.
\eee
The identification (\ref{gdef}) tells us the transformation properties
of the Lagrange multipliers. Note that for $p=0,1$ the auxiliary
metric $G_{ij}$ never appears in (\ref{Gact}), and the configuration
space action is the usual manifestly reparametrization invariant
massive point-particle action and Brink-Howe-DiVecchia-Deser-Zumino
reparametrization invariant tensile string action \cite{BRIN,DESE},
respectively.

It is clear from the above procedure that we may take the limit $T\to
0$ anywhere between (\ref{hami}) and (\ref{gdef}). The identification
(\ref{gdef}) will differ in that limit, however. The metric density
$T\sqrt{-g}g^{\alpha \beta}$ becomes degenerate and gets replaced by a
rank 1 matrix which can be written as $V^\alpha V^\beta$ in terms of
the vector density $V^\alpha$
\be
V^\alpha \leftrightarrow {1 \over {\sqrt{2}\lambda}}(1,\rho ^a).
\ee
In fact, using this prescription the $T\to 0$ limit of the $p$-brane
action is
\be
S=\int{d^{p+1}\xi V^\alpha V^\beta\partial _\alpha X^m \partial _\beta
X^n \eta _{mn}}.
\ee

We will henceforth be concerned with the string case
\bbe{nullact}
S=\int{d^2\xi V^\alpha V^\beta\partial _\alpha X^m \partial _\beta X^n
\eta _{mn}}
\eee
where $V^\alpha$ are world sheet vector densities of weight $1\over 4$
and the world sheet coordinates are $\xi ^\alpha
\equiv (\tau ,\sigma)$.
There exists many formulations of the bosonic tensionless string.  The
present has many advantages due to its geometric form. First, in the
quantum theory in going from the phase space action (\ref{phact}) to
the configuration space action (\ref{cact}) one generates a functional
determinant which may be interpreted as a modification of the path
integral measure. This modification is precisely what is needed to
render the measure invariant under 2D diffeomorphisms, i.e., it leads
to the Fujikawa variables. This is all the more important since those
variables are normally defined by weighting with various powers of the
determinant of the 2D metric and here we have no such determinant
available. Second, the action (\ref{nullact}) is easy to
supersymmetrize \cite{ulbsgt1}. In contrast, the action derived as in
(\ref{phiact}), e.g.;
\be
I_1^0=\int {d^2\xi \phi \det }\gamma _{\alpha\beta}
\ee
cannot be easily extended to the superstring because of the Siegel
symmetry of the superstring which transforms the Lagrangian $\sqrt
{-det \gamma_{\alpha\beta}}$ into a 2D total derivative. In general,
we see from the relation between the actions (\ref{act1}) and
(\ref{phiact}) that whereas $\delta {\cal L}=\partial\cdot\omega$
leaves the action (\ref{act1}) invariant for some transformation with
parameter $\omega$, for invariance of the action (\ref{phiact}) in the
limit $T\to 0$, we need a transformation $\delta\phi =-\phi {\cal
L}^{-1}\partial\cdot\omega$. Clearly this leads to difficulties at the
level of field equations where ${\cal L}=0$. Third, when introducing
spin via world sheet supersymmetry it leads naturally to a new 2D
superspace geometry which in turn allows for a compact treatment of
many different classes of models \cite{roli}.

\bigskip
\begin{flushleft}
{\bf 2.2 Symmetries}
\end{flushleft}
\bigskip

The action (\ref{nullact}) for the tensionless string is invariant
under world-sheet diffeomorphisms and space-time conformal
transformations. Under the diffeomorphisms $X^m$ transforms as a
scalar field
\bbe{a}
\delta _\varepsilon X^m = \varepsilon \cdot \partial X^m,
\eee
and $V^\alpha$ as a vector density:
\bbe{b}
\delta _\varepsilon V^\alpha =-V \cdot \partial \varepsilon ^\alpha
+\varepsilon \cdot \partial V^\alpha +{\textstyle{1 \over 2}}(\partial
\cdot
\varepsilon )V^\alpha .
\eee

There are of course many different gauge choices possible for fixing
the reparametrization symmetry (\ref{b}). We have found the following
{\it transverse gauge} particularly useful:
\bbe{g}
V^\alpha = (v,0),
\eee
with $v$ a constant. The transverse gauge corresponds to the conformal
gauge $g_{\alpha
\beta}= e^\phi \eta _{\alpha \beta}$ in the tensile theory. For classical
string
propagation, where the world-sheet is cylindrical, one can always
reach this gauge, except in the particular case when field lines of
the vector density are closed around the cylinder. In this exceptional
case, which we do not consider here, one may instead choose $V^\alpha
= (0,v)$. For a toroidal geometry there is a continuum of globally
inequivalent vector densities. The physical consequences of this fact
are currently under investigation.

Just as in the tensile case there is a residual symmetry that leaves
(\ref{g}) invariant;
\bbe{h}
\delta \xi ^\alpha = \lambda ^\alpha, \qquad \lambda ^\alpha = (f'(\sigma
)\tau + g(\sigma ),f(\sigma ))
\eee
with $f$ and $g$ arbitrary functions of $\sigma $ only\footnote{Note
that although we give the infinitesimal form of the transformations in
(\ref{h}), the finite form looks the same with $\xi ^\alpha =\lambda
^\alpha$ and different functions $f$ and $g$.}. If we define the
generators
\bbe{RTgen}
R(f)&\equiv &f'(\sigma)\tau \partial_\tau +f(\sigma)\partial_\sigma
\cr T(g)&\equiv &g(\sigma)\partial_\tau ,
\eee
we find the following algebra
\bbe{RTalg}
\left[{R(f_1),R(f_2)}\right]&=&R(f_1f'_2-f_2f'_1),\cr
 \left[{R(f),T(g)}\right]&=&T(fg'-gf'),\cr
\left[{T(g_1),T(g_2)}\right]&=&0
\eee
If we furthermore fourier expand $f$
\be
f(\sigma )={\textstyle{1 \over {2\pi}}}\sum\limits_n {a^f_ne^{2\pi
in\sigma}},
\ee
we may write
\bbe{neweq}
R(f)&=&{\textstyle{1 \over {2\pi}}}
\sum\limits_n {a^f_ne^{2\pi in\sigma}\left({2\pi in\tau \partial_\tau
+\partial_\sigma}\right)}\equiv -i\sum\limits_n {a^f_n R_n}\cr T(g)&=&
{\textstyle{1 \over {2\pi}}}
\sum\limits_n {a^g_ne^{2\pi in\sigma} \partial_\tau}\equiv
-i\sum\limits_n {a^g_n T_n}.
\eee
Despite their somewhat complicated form the generators $R^m$ then
satisfiy the algebra of infinitesimal diffeomorphisms of $S^1$, i.e.,
the Virasoro algebra:
\be
\left[{R_m,R_n}\right]=(m-n)R_{m+n} + {\textstyle{1 \over
12}}(C_R-C_L)(m^3-m)\delta_{m+n}
\ee
and for the mixed commutator we find
\be
\left[{R_m,T_n}\right]=(m-n)T_{m+n}+{\textstyle{1 \over
12}}(C_R+C_L)(m^3-m)\delta_{m+n}.
\ee
Here we have also displayed the most general central extension
compatible with the Jacobi identities. A simple $T\to 0$ limit of the
tensile string algebra yields zero central charges, i.e., $C_L=C_R \to
0$, (but other models are not logically excluded).  Thus the residual
symmetry (\ref{h}) is the semi-direct product of $\sigma$-dependent
$\tau$-translations with a Virasoro symmetry.

The Poincar\'e symmetry is extended to conformal symmetry for massless
particles and massless free fields. The conformal group preserves the
causal structure of Minkowski space and maps light cones onto light
cones. We shall see that in the transverse gauge the tensionless
string can be viewed as a collection of massless particles and it is
thus natural to expect that classical tensionless strings should enjoy
conformal symmetry. In fact, since a conformal transformation will
scale the D-dimensional line element, it will also scale the induced
metric. This can be compensated by a ($X^m$-dependent) scaling of
$V^\alpha$, and the action (\ref{nullact}) thus be left invariant (see
(\ref{c}) below).  Note that this is not possible for the action of
the tensile string,
\bbe{d} S_1=-{T \over 2}\int {d^2}\xi \sqrt
{-\det g_{\gamma\delta}}g^{\alpha\beta}\gamma _{\alpha\beta},
\eee
since any rescaling of $g_{\alpha\beta}$ is an invariance of $\sqrt
{-det g_{\gamma\delta}}g^{\alpha\beta}$ alone. In fact, in this sense
{\it world-sheet Weyl-invariance is replaced by space-time conformal
invariance} in the limit $T \to 0$.

The infinitesimal transformations form the conformal algebra:
\bbe{c.alg}
\lbrack m_{mn},m^{rs}] &=& \delta_m^r m_n^{~s} - \delta_n^r m_m^{~s} +
\delta_n^s
m_m^{~r} - \delta_m^s m_n^{~r} \cr
\lbrack m_{mn},p^s] &=& -\delta_n^s p_m +\delta_m^s p_n, \qquad
\lbrack p_m,p_n] = 0 \cr
\lbrack m^{mn},k_l] &=& -\delta^n_l k^m + \delta^m_l k^n, \qquad
\lbrack k^m,k^n] = 0\cr
\lbrack p_m,k^n] &=& -\delta_m^n s +m_m^{~n}\\
\lbrack s, m_{mn}] &=& 0\cr
\lbrack s,p_m] &=& p_m\cr
\lbrack s,k^m] &=&  -k^m \nonumber
\eee
Here $m_{mn},p_m,k^n$ and $s$ are the generators of Lorentz
transformations, translations, conformal boosts and dilatations,
repectively.  Note that the whole algebra can be generated from
repeated brackets of $p_m$ and $k^n$.

Under Poincar\'e transformations $X^m$ behaves as a Lorentz vector and
$V^\alpha$ as a scalar. In contrast, conformal boosts (generator
$k^m$) and dilations (generator $s$) rescale $V^\alpha$ in order to
leave the action invariant when the induced metric (\ref{indm}) is
rescaled by the ordinary action of conformal transformations on the
coordinates $X^m$. The infinitesimal conformal boosts and dilatations
act as follows:
\bbe{c}
\delta _b X^m &=& [b\cdot k,X^n]=(b \cdot
X)X^m-\half X^2b^m, \cr
\delta _b V^\alpha &=&-b \cdot XV^\alpha\cr
\delta _a X^m&=&[as,X^m]=aX^m, \cr
\delta _a V^\alpha&=& -aV^\alpha
\eee
where $b_m$ and $a$ are transformation parameters.  The finite form of
the special conformal transformation of $V^\alpha$ reads
\bbe{FinV}
V'^\alpha =V^\alpha\sqrt{1+2b\cdot X+b^2X^2},
\eee
which shows that conformal transformations may take $V^\alpha$ to zero
for some string solutions. Therefore the transverse gauge cannot be
imposed globally on the world-sheet for such string states. The
connection, if any, of this fact to the local problem of the
infinitesimal quantum transformations described below remains to be
elucidated.

Let us finally mention in passing that there exists a formulation of
the zero tension string where the conformal transformations act
linearly. This is achieved by describing the string in a target space
with one additional spacelike and one additional timelike coordinate
\cite{akul}. Recently a hamiltonian treatment of this formulation was
given \cite{ISBE}. The symmetries (\ref{h}) are then enlarged to a
particular semi-direct sum between an $SU(1,1)$ affine Ka\v c-Moody
and a Virasoro algebra.  Essentially, the two additional symmetry
generators are needed to compensate for the extra dimensions.

\bigskip
\begin{flushleft}
{\bf 2.3 Equations of motion}
\end{flushleft}
\bigskip

The field equations that follow from the action (\ref{nullact}) are:
\bbe{e}
V^\beta \gamma _{\alpha \beta}=0,\qquad \partial _\alpha (V^\alpha
V^\beta \partial _\beta X^m )=0
\eee
The first of these equations states that $\gamma_{\alpha\beta}$ has an
eigenvector with eigenvalue zero which implies that it is a degenerate
matrix:
\bbe{f}
\det \gamma _{\alpha \beta}=0
\eee
This means that the world sheet spanned by the tensionless string is a
{\it null surface}. For this reason tensionless strings are sometimes
referred to as "null strings".

The second of the field equations is most easily interpreted in the
gauge (\ref{g}).  In this gauge the equations (\ref{e}) become
\bbe{eq.mo}
\ddot X^m &=& 0 \cr
\dot X^2 &=& \dot X \cdot X'=0.
\eee
Clearly the string behaves classically as a collection of massless
particles, one at each $\sigma$ position, constrained to move
transversally to the direction of the string.

For open strings there are also edge conditions. With $\sigma \in
\left[{0,1}\right]$, we find from the derivation of (\ref{e}), that
we need to demand
\bbe{edge}
\left[{V^1V^\alpha \partial_\alpha X_m\delta
X^m}\right]_{\sigma=0,1}=0.
\eee
We may implement this by requiring either
\bbe{cnd1}
V^1(\tau,0)=V^1(\tau,1)=0,
\eee
or
\bbe{cnd2}
V^\alpha\partial_\alpha X^m(\tau,0)=V^\alpha\partial_\alpha
X^m(\tau,1)=0.
\eee
The condition (\ref{edge}) can be satisfied simply by choosing a gauge
which approaches the transverse gauge (\ref{g}) at the edges of the
string, thus fulfilling (\ref{cnd1}).  If we want to be able to impose
a "non-transverse" gauge where $V^1\not = 0$, (\ref{cnd2}) has to be
satisfied. For example, in the gauge $V^\alpha = (0,v)$ it happens to
yield the usual $T \not =0$ open string edge conditions
\bbe{Tedge}
X'^m(\tau,0)=X'^m(\tau,1)=0
\eee

The first equation in (\ref{e}) is the $V^\alpha$-field equation. It
corresponds to the $g_{\alpha\beta}$-equation $T_{\alpha\beta}=0$ in
the tensile theory, i.e., to the Virasoro constraints. In our case the
energy-momentum tensor cannot be derived as a field-equation in the
same way, but it is nevertheless related to the $V^\alpha$-field
equation. The energy momentum mixed tensor(density) with one covariant
and one contravariant index is derived as the translation current. It
reads
\bbe{enmo} T_\alpha^\beta
=V^\beta V^\sigma \partial_\sigma X^m\partial_\alpha X_m -\half
V^\sigma V^\rho \partial_\sigma X^m \partial_\rho
X_m\delta_\alpha^\beta .
\eee
It is traceless and comparing to (\ref{e}) we see that it vanishes on
$V^\alpha$-shell.  Furthermore, in analogy to the tensile case, we
expect it to be covariantly conserved using the second equation in
(\ref{e}) only.  To discuss covariant conservation one has to add more
geometric structure to the theory than is needed in the action
principle. Introducing a covariant derivative as in \cite{ulbsgt2} or
\cite{roli}, we find that \be
\nabla_\alpha T_\beta^\alpha =0 \ee
provided that
\be
0=\nabla_\alpha V^\beta \equiv \partial_\alpha V^\beta +
\Gamma_{\alpha\sigma}^\beta V^\sigma  -\half \Gamma _{\alpha
\sigma}^\sigma V^\beta,
\ee
the analogue of the metricity condition in the non-degenerate case.
It is interesting that only the weaker condition
\be
\nabla_\alpha V^\alpha =0
\ee
was needed in \cite{ulbsgt2} and \cite{roli}. We shall find no more
use for the connection $\Gamma_{\alpha\sigma}^\beta$ in this article.

In the transverse gauge (\ref{g}) the components of the
energy-momentum tensor (\ref{enmo}) are
\bbe{zenm}
T_0^0=-T_1^1= \half v^2\gamma_{00}, \qquad T_1^0=v^2\gamma_{10}
\qquad T_0^1=0.
\eee
Since there is no metric to raise and lower indices, we cannot ascribe
the usual symmetry properties to $T_\alpha^\beta$. Nevertheless it
still has only two independent components. Using Poisson brackets the
components in (\ref{zenm}) is readily seen to generate the semi-direct
product of $\sigma$-dependent $\tau$-translations and a Virasoro
symmetry discussed in the previous section. Thus the energy momentum
tensor generates a symmetry which becomes the residual symmetry
(\ref{RTalg}) after gauge fixing to transversal gauge, in complete
analogy to the tensile case.

\bigskip
\begin{flushleft}
{\bf 2.4 The light-cone gauge}
\end{flushleft}
\bigskip

We introduce light cone coordinates $(X^+ ,X^- ,X^i)$, where $X^\pm
\equiv {1
\over {\sqrt 2}}(X^0\pm X^{D-1})$ and $i=1...D-2$. This is
only a choice of coordinates, but next we use the residual symmetry
(\ref{h}), and the equation of motion (\ref{eq.mo}), to fix a light
cone gauge. From
\bbe{j}
\tilde \tau =F'(\sigma )\tau +G(\sigma), \qquad \ddot X ^+=0
\eee
we see that we may take $\tilde \tau \propto X^+$, and we choose
\bbe{k}
X^+=\frac{p^+}{v^2}\tau
\eee
where $p^+$ is the conserved momentum. This completely fixes the
diffeomorphism gauge, except for rigid $\sigma$-translations.

In light cone coordinates the $V^\alpha$ equations of motion read
\bbe{m}
V^\alpha \partial _\alpha X^i \partial _\beta X^i -V^\alpha \partial
_\alpha X^- \partial _\beta X^+ -V^\alpha \partial _\alpha X^+
\partial _\beta X^- = 0 ,
\eee
and in transverse gauge (\ref{g}) they give the constraints
\bbe{n}
\dot X^i \dot X^i -2\dot X^- \dot X^+ &=& 0\cr
\dot X^i X'^i -\dot X^- X'^+-\dot X^+ X'^- &=& 0.
\eee

We now use (\ref{n}) in the light cone gauge (\ref{k}) to eliminate
$X^-$, except for a zero-mode $x^-(\tau)$:
\bbe{gnutt}
X'^- &=& {v^2 \over p ^+}\dot X^i X'^i\cr
\dot X^- &=& {v^2 \over 2p ^+}\dot X^i \dot X^i \cr
x^- &\equiv& \int d\sigma X^- \equiv x^-_0 + \frac{1}{v^2}p^-_0\tau \;
,
\eee
where $x^-_0$ and $p^-_0$ are constants.  Having eliminated $X^\pm $
we are left with the equations of motion for the transverse
components, $X^i$, (in transverse gauge)
\bbe{p}
 \ddot X^i=0
\eee
These may be derived from the light cone action
\bbe{q}
S_{LC}=\frac{v^2}{2}\int {d^2\xi \dot X^i \dot X^i}
\eee

We now wish to find a canonical formulation of the generators in order
to prepare for quantization.  The transverse conjugate momenta can be
read off from (\ref{q}), they are
\bbe{r}
P ^i=v^2 \dot X^i
\eee
In addition $-p^+$ is canonically conjugate to $x^-$.  We now write
the first two equations in (\ref{gnutt}) using (\ref{r})
\bbe{mx}
X'^-={1 \over p ^+ } X'^i P^i
\eee
\bbe{s}
\dot X^-={1 \over 2p ^+ v^2}P ^i P ^i \equiv \frac{1}{v^2}P^-
\eee
The action (\ref{q}) then corresponds to a Hamiltonian
\bbe{LCH}
{\cal H}={\textstyle{1 \over {2v^2}}}\int{d\sigma P^i
P^i}={\textstyle{1 \over {v^2}}}\int{d\sigma p ^+ P ^-},
\eee
which indeed generates $\tau$-translations, c.f. (\ref{k},\ref{gus}).
Equations (\ref{k},\ref{r}) and (\ref{s}) give the translation
operators:
\bbe{gus}
p^m \equiv \int d\sigma P^m(\sigma) .
\eee
Here we have introduced the convention that lower case letters denote
zero modes or integrated quantities.

The generators of the conformal group additional to the Lorentz and
translation generators, can now be written (at $\tau =0$):
\bbe{x}
\!\!\!p^i &=& \int {d\sigma P^i}\cr
\!\!\!p^- &=& \int {d\sigma P^-}\cr
\!\!\!p^+ &=& p^+\cr
\!\!\!m^{ij} &=& \int {d\sigma \left\{{X^i P^j
-X^j P^i }\right\}}\cr
\!\!\!m^{i-} &=& \int {d\sigma \left\{{X^i P^-
-X^- P^i }\right\}}\cr
\!\!\!m^{i+} &=& \int {d\sigma \left\{{p^+X^i }\right\}}=p^+x^i\cr
\!\!\!m^{+-} &=& -\int {d\sigma \left\{{p^+X^- }\right\}}=-p^+x^-\cr
\!\!\!s &=& \int {d\sigma \left\{ {X^i P ^i -
p^ + X^- } \right\}}\cr
\!\!\!k^i &=& \int {d\sigma \left\{X^i X^j P^j
-X^i p^ + X^- -\half X^jX^jP^i \right\}}\cr
\!\!\!k^- &=&\! {1\over{p ^+}}\int{
\!d\sigma \left\{{ p ^+X^j
X^- P ^j }\right.  }\!\!\cr && -\left.{ p ^+X^- p ^+X^-
\!-\!\half p^+
X^i X^i P ^- }\right\}\cr
\!\!\!k^+ &=&-\half \int{d\sigma \left\{p ^+X^i X^i \right\}}
\eee
The generators in (\ref{x}) generate transformations that include
precisely the compensating reparametrizations derived in the next
section.

\bfl
{\bf 2.5 Compensating reparametrizations}
\efl\jump

In this section we wish to find the precise form of the combined
special conformal transformations and diffeomorphisms that preserve
both the choice of transverse gauge and light cone gauge. They may be
found following the procedure used by Goddard et al.
\cite{ggrt} for the Lorentz transformations of the ordinary string.

Under special conformal transformations the fields in the action
(\ref{nullact}) transform as (c.f. (\ref{c})):
\bbe{sctf}
  \delta_b X^m &=& b\cdot\!  X X^m -\half X^2 b^m\cr \delta_b
\partial_\alpha X^m &=& b\cdot\!  (\partial_\alpha X) X^m + b\cdot\!
X \partial_\alpha X^m - X\cdot\!  \partial_\alpha b^m\\ \delta_b
V^\alpha &=& -b\cdot\!  X V^\alpha\cr \delta_b \gamma_{\alpha\beta}
&=& 2 b\cdot\!  X \gamma_{\alpha\beta}\nonumber
\eee
where $b^m$ is a constant vector. The transformation rule for
$V^\alpha$ follows from demanding invariance of (\ref{nullact}).

 Under infinitesimal reparametrizations $\xi^\alpha \rightarrow
\xi^\alpha - \varepsilon^\alpha(\xi^\beta)$ the field transformations
are (c.f.  (\ref{a},\ref{b})):
\be
  \delta_\varepsilon X^m &=& \varepsilon^\delta\partial_\delta X^m\cr
\delta_\varepsilon \partial_\alpha X^m &=&
\varepsilon^\delta\partial_\delta\partial_\alpha X^m + (\partial_\beta
X^m) \partial_\alpha\varepsilon^\beta\\ \delta_\varepsilon V^\alpha
&=& \varepsilon^\delta\partial_\delta V^\alpha-
V^\delta\partial_\delta\varepsilon^\alpha +
\half(\partial_\delta\varepsilon^\delta)V^\alpha\cr \delta_\varepsilon
\gamma_{\alpha\beta} &=&
\varepsilon^\delta\partial_\delta\gamma_{\alpha\beta} +
(\partial_\alpha\varepsilon^\gamma)\gamma_{\alpha\beta}+
(\partial_\beta\varepsilon^\gamma)\gamma_{\alpha\gamma}\nonumber
\ee
\ie they transform as a scalar, covariant vector, contravariant vector density
and contravariant tensor respectively.

Let $\delta^{l.c.}_{b}$ denote the combined actions of
reparametrizations and special conformal transformations \ie
$\delta^{l.c.}_{b}=\delta_\varepsilon + \delta_b$.  In order to stay
in the transverse gauge (\ref{g}) we must have
\bbe{e1}
 0 &=& \delta^{l.c.}_{b}V^0 = -b\cdot\!  X v - v \dot{\varepsilon}^0
+\half v \dot{\varepsilon}^0 +\half v \varepsilon'^1\\ 0 &=&
\delta^{l.c.}_{b}V^1 =-v \dot{\varepsilon}^1\nonumber
\eee
where dot (prime) refers to $\tau=\sigma^0$ $(\sigma=\xi^1)$
derivatives.  This implies
\bbe{e2}
  && \dot{\varepsilon}^0 - \varepsilon'^1 = -2b\cdot\!  X \\ &&
\dot{\varepsilon}^1 =0\nonumber
\eee
and in particular we see that $\varepsilon^1$ depends on $\sigma$
only.  Note also that as long as (\ref{g}) is the only gauge condition
imposed we can choose a compensating reparametrization in several ways
for any given special conformal transformation.  However, we also wish
to stay in the light cone gauge, which we choose to define as
\bbe{newformula}
X^+(\tau,\sigma)=\dot{X}^+(\tau,\sigma)\:\tau,\quad
\partial_\alpha\dot{X}^+(\tau,\sigma)=0.
\eee
This formulation, using a Lagrangian language rather than a
Hamiltonian one, is useful since it makes manifest the transformation
properties of all factors under reparametrizations. To stay in this
gauge requires further restrictions on $\varepsilon^\mu$ as we will
now see.

For our purposes it suffices to study transformations on shell. In the
transverse gauge (\ref{g}) we have the equations of motion and
constraints (\ref{eq.mo}) which are solved by
\bbe{soln}
  X^m(\tau,\sigma) = \dot{X}_0^m(\sigma)\: \tau +X^m_0(\sigma)\quad
{\rm if}\quad \dot{X}_0^2(\sigma)=\dot{X}_0(\sigma)\cdot\!
X'_0(\sigma) =0
\eee
where $\dot{X}_0^m(\sigma)\equiv\dot{X}^m(0,\sigma)$ and $
X_0^m(\sigma)\equiv X^m(0,\sigma)$ are initial values.  The
requirement of the invariance of (\ref{newformula}) reads
\be
 0&=& \delta^{l.c.}_{b}(X^+ -\dot{X}^+\tau)\; =\;
\delta^{l.c.}_{b}X^+-(\delta^{l.c.}_{b}\dot{X}^+)\tau = \cr &=&
b\cdot\!  X X^+ -\half b^+ X^2 + \varepsilon^0\dot{X}^+ -\\ && -
(b\cdot\!  \dot{X} X^+ +b\cdot\!  X \dot{X}^+ - b^+ X\cdot\! \dot{X}
+\dot{\varepsilon}^0 \dot{X}^+)\tau\nonumber
\ee
where we have used (\ref{e1},\ref{e2},\ref{eq.mo}). Using (\ref{soln})
then gives
\bbe{Xyzzy}
  b\cdot\!  \dot{X}_0 \dot{X}_0^+ \tau^2 +\half b^+X_0^2 = \dot{X}_0^+
(\varepsilon^0 -
\dot{\varepsilon}^0 \tau ).
\eee
To solve this we make the ansatz
\be
\varepsilon^0(\tau,\sigma) = r_1(\sigma) + r_2(\sigma)\tau +r_3(\sigma)\tau^2
\ee
 with coefficient functions $r_1,r_2$ and $r_3$, to be determined.
Inserting this into (\ref{Xyzzy}) immediately gives us
\bbe{ee1}
  \varepsilon^0 =\frac{1}{2\dot{X}_0^+} b^+ X_0^2 + r_2\tau - b\cdot\!
\dot{X}_0\tau^2
\eee
where $r_2$ is still undetermined. This solution for $\varepsilon^0$,
as can be easily checked, is also consistent with
$\ddot{\varepsilon}^{\:0} = -2b\cdot\! \dot{X}$ which follows from
(\ref{e2}). Inserting this solution into (\ref{e2}) gives
\bbe{ee2}
\varepsilon'^1 &=& \dot{\varepsilon}^0 + 2b\cdot\!  X = -2b\cdot\!
\dot{X}_0\tau
+r_2+ 2b\cdot\!  X = r_2 + 2b\cdot\!  X_0\\
\varepsilon^1 &=& 2b\cdot\! \int_o^\sigma X_0 d\sigma'
   +\int_o^\sigma r_2 d\sigma' + {\rm const.}\nonumber
\eee

For consistency we also have to insure that
$\partial_\alpha\dot{X}^+_0 = 0$ is invariant under the
transformation. That is, we have to check whether $\delta^{l.c.}_{b}
\ddot{X}^+$ and $\delta^{l.c.}_{b}\dot{X}'^+$ vanish on-shell. The
first of these can be shown to vanish identically and yields nothing
new, while the latter condition will determine the coefficient $r_2$.

To accomplish this we first note that the second derivative of a field
$\phi$ does not transform as a tensor but has the transformation law
\be
  \delta_\varepsilon \partial_\alpha \partial_\beta \phi =
\varepsilon^\gamma\partial_\gamma \partial_\alpha \partial_\beta \phi
+ (\partial_\alpha\varepsilon^\gamma) \partial_\gamma \partial_\beta
\phi + (\partial_\beta\varepsilon^\gamma) \partial_\gamma
\partial_\alpha \phi +
(\partial_\alpha\partial_\beta\varepsilon^\gamma) \partial_\gamma \phi
\ee
under reparametrizations. In the light-cone gauge this takes a
particularly simple form for the transformations of
$\partial_\alpha\dot{X}^+$:
\be
  \delta_\varepsilon \partial_\alpha \partial_0 X^+ &=&
(\partial_\alpha \dot{\varepsilon}^0) \dot{X}^+\\ \delta_b
\partial_\alpha \partial_0 X^+ &=& \partial_\alpha (b\cdot\!
\dot{X}X^+ + b\cdot\!  X \dot{X}^+ - X\cdot\!  \dot{X} b^+ ).\nonumber
\ee
Using this we obtain the combined on-shell transformation properties:
\bbe{gugg}
 \delta^{l.c.}_{b} \ddot{X}^+ &=& b\cdot\! \dot{X} \dot{X}^+ +
b\cdot\! \dot{X} \dot{X}^+ + \ddot{\varepsilon}^{\: 0}\dot{X}^+ = 0\cr
&&\cr \delta^{l.c.}_{b} \dot{X}'^+ &=& b\cdot\! \dot{X}' X^+ +
b\cdot\!  X' \dot{X}^+ - X\cdot\! \dot{X}' b^+ +
\dot{\varepsilon}'^0\dot{X}^+ =\\ &=& b\cdot\! \dot{X}' X^+ + b\cdot\!
X' \dot{X}^+ - X\cdot\! \dot{X}' b^+ - 2 b\cdot\! \dot{X}_0'\dot{X}^+
\tau +s'\dot{X}^+ =\cr &=& b\cdot\!  X'_0 \dot{X}_0^+ - X_0\cdot\!
\dot{X}_0' b^+ + r'_2\dot{X}_0^+.\nonumber
\eee
The first of these vanish identically as promised and for the second
to vanish we get
\be
  r'_2= X_0\cdot\! \dot{X}_0' \frac{b^+}{\dot{X}_0^+} - b\cdot\!
X'_0.
\ee
This can be integrated explicitly since $\dot{X}_0 \cdot\!  X'_0$
vanishes:
\be
   r_2 = X_0\cdot\! \dot{X}_0 \frac{b^+}{\dot{X}_0^+} - b\cdot\!  X_0
+ C_1
\ee
where $C_1$ is some arbitrary integration constant. Plugging this
value of $r_2$ into (\ref{ee2}) then yields
\be
\varepsilon^1(\sigma) =\int_0^\sigma d\sigma' \left(b\cdot\!  X_0(\sigma') +
    X_0(\sigma')\cdot\! \dot{X}_0(\sigma')
\frac{b^+}{\dot{X}_0^+}\right) + C_1\sigma + C_2.
\ee
Since we are dealing with the closed string the periodicity condition
$\varepsilon^1(0)=\varepsilon^1(1)$ must be satisfied, and this can be
used to determine $C_1$:
\be
\varepsilon^1 &=&\int_0^\sigma\!\! d\sigma' \left(b\cdot\!  X_0 +
    X_0\cdot\! \dot{X}_0\: \frac{b^+}{\dot{X}_0^+}\right) \cr
&-&\sigma\!\!\int_0^1\!\! d\sigma' \left(b\cdot\!  X_0 + X_0\cdot\!
\dot{X}_0\: \frac{b^+}{\dot{X}_0^+}\right) + C_2.
\ee
This, however, can be written in a simpler and manifestly periodic
form
\be
\varepsilon^1(\sigma) =\oint d\sigma' \left(b\cdot\!  X_0(\sigma') +
    X_0(\sigma')\cdot\! \dot{X}_0(\sigma')
\:\frac{b^+}{\dot{X}_0^+}\right) h(\sigma'-\sigma) +C_2
\ee
where
\bbe{ad}
h(\sigma - \tilde \sigma) \equiv \sigma -\tilde \sigma -{\textstyle{1
\over 2}{\rm sign} (\sigma -\tilde \sigma ) } .
\eee
Thus we conclude that the special conformal transformations plus
compensating reparametrizations that preserve both the transverse and
the light cone gauge act as follows:
\be
 \delta^{l.c.}_{b}X^m(\tau,\sigma) &=& b\cdot\!  X(\tau,\sigma)
X_m(\tau,\sigma) -\half X^2(\tau,\sigma) b_m \;\;+\cr &&
+\;\;\varepsilon^0(\tau,\sigma)\dot{X}^m(\tau,\sigma)
+\varepsilon^1(\sigma) X^{'m}(\tau,\sigma)\\ &&\cr
\varepsilon^0(\tau,\sigma) &=& \frac{1}{2\dot{X}_0^+} b^+
X_0^2(\sigma)+ r_2(\sigma)\tau - b\cdot\!  \dot{X}_0 (\sigma)\tau^2\cr
r_2 (\sigma) &=& X_0 (\sigma)\cdot\! \dot{X}_0 (\sigma)
\frac{b^+}{\dot{X}_0^+} - b\cdot\!  X_0 (\sigma) -\cr && - \oint
d\sigma' \left(b\cdot\!  X_0(\sigma') + X_0(\sigma')\cdot\!
\dot{X}_0(\sigma') \frac{b^+}{\dot{X}_0^+}\right)\cr
\varepsilon^1(\sigma) &=&\oint d\sigma' \left(b\cdot\!  X_0(\sigma') +
X_0(\sigma') \cdot\! \dot{X}_0(\sigma') \frac{b^+}{\dot{X}_0^+}\right)
h(\sigma'-\sigma) +C_2 .\nonumber
\ee
The constant $C_2$ is not fixed by these considerations. However, we
recall that gauge fixing to the light-cone gauge leaves a rigid
$\sigma$-translation unspecified. The $C_2$-term generates precisely
this remaining gauge symmetry.

\begin{flushleft}
\section{The quantum theory}
\end{flushleft}

It is only possible to consistently quantize the tensile string in
certain critical dimensions. This result can be arrived at in a number
of different ways: By demanding that the Weyl invariance holds at the
quantum level (absence of the conformal anomaly); by demanding
Lorentz-symmetry in the light cone gauge or by demanding nilpotency of
the BRST charge. For the tensile bosonic string all these methods (and
a few other) lead to the critical dimension $D$=26.

The question of whether the quantization of the tensionless string
leads to similar restrictions may likewise be investigated via various
routes. The first of the above alternatives is not available, though,
since the action has no Weyl symmetry. In the main part of the
remainder of this article we shall be concerned with light-cone gauge
quantization and the consequences of requiring the full space-time
symmetry of the classical model (the conformal symmetry) to be
preserved by quantization. We will also confirm the known result that
the weaker requirement of quantum Poincar\'e \ symmetry does not lead
to any quantization problems. However, we first comment on
BRST-quantization.

\bigskip
{\bf 3.1 BRST-Quantization}:\\
\bigskip

BRST-quantization may be applied either in a Hamiltonian or in a
Lagrangian context.

A Hamiltonian BRST-quantization of the bosonic tensionless string was
carried out in
\cite{liraspsr}. In phase space the second pair of equations in (\ref{eq.mo})
read
\bbe{constraints}
 P^2=P\cdot X'=0
\eee
Starting from the algebra of these constraints the authors of
\cite{liraspsr} construct the Hamiltonian BRST-charge $Q_H$, following
the procedure used in, e.g., \cite{HWAN}.  Nilpotency of $Q_H$ is then
checked in the quantized theory and found to hold independent of the
dimension. This procedure says nothing about the space-time conformal
symmetry, of course. It has been extended to include this symmetry in
a mode-expansion approach in \cite{ISBE}. There obstructions to
quantization are found.

As an alternative, we here consider the Lagrangian BRST quantization,
following
\cite{KATO}. The 2D diffeomorphisms transformations of the fields
are given by (\ref{a},\ref{b}) where $\varepsilon ^\alpha$ is the
transformation parameter. Following the standard BRST procedure we
then introduce anticommuting ghosts $c ^\alpha$, antighosts $\bar c
_\alpha$ and auxiliary fields $B _\alpha$. The BRST transformations
are

\bbe{BRStf}
sX^m =c\partial X^m ,\qquad sV^\alpha =c\partial V^\alpha -V\partial
c^\alpha +\half V^\alpha\partial c\cr sc^\alpha =-c\partial
c^\alpha,\qquad s\bar c_\alpha =iB_\alpha,\qquad sB _\alpha =0
\eee
The gauge fixing fermion that implements the transverse gauge
(\ref{g}) is
\bbe{GFF}
\Psi =-i\left( {\bar c_0(V^0-v)+\bar c_1V^1} \right)
\eee
with $v$ a constant. The gauge fixing and ghost Lagrangian is obtained
as $s\Psi$. The total gauge fixed action is
\bbe{GFA}
S_{GF}=S_1^0+\int {d^2}\xi s\Psi
\eee
Written out in detail it is rather complicated, but after a
redefinition of the auxiliary fields, $B \to \hat B$,
\bbe{BHAT}
\hat B_\alpha \equiv B_\alpha &+&i\bar c_\beta \partial_\alpha c^\beta -\ihalf
\bar
c_\alpha (\partial\cdot c) \cr &+& V_\alpha + i\partial_\beta (\bar
c_\alpha c^\beta ) + V^\beta \gamma_{\beta
\alpha}
\eee
it simplifies to
\bbe{SGFS}
S_{GF}=\int {d^2}\xi \left[ {v^2\partial _0X^m\partial_0 X_m+\hat
B_0(V^0-v)+\hat B_1V^1+i\bar c_\alpha D_\beta^\alpha c^\beta} \right]
\eee
where
\bbe{DDDD}
D_\beta^\alpha=v\left( \matrix{{\textstyle{1 \over 2}}\partial _0\quad
-{\textstyle{1 \over 2}}\partial _1\hfill\cr 0\qquad \quad \;\partial
_0\hfill\cr} \right).
\eee
{}From this action we derive the momenta
\bbe{MOMENT}
\Pi _m =2v^2\dot X_m ,\qquad \Pi _0^c={{iv} \over 2}\bar c_0,\qquad \Pi
_1^c=iv\bar c_1,
\eee
conjugate to $X ^m, c ^0$ and $c ^1$, and all other momenta vanish.
The Hamiltonian we find is
\bbe{HAMI}
H=\int{d\sigma \left[{ {1\over{4v^2}}\Pi _m \Pi ^m + \Pi
_0^cc'^1}\right]}
\eee
and the BRST charge is
\bbe{LBRS}
Q=\int {d\sigma \left[ {-{1 \over {4v^2}}c^0\Pi _m \Pi ^m -\Pi _m X'^m
c^1+\Pi _0^cc\cdot\partial c^0+\Pi _1^cc\cdot\partial c^1} \right]}
\eee
One may check that $Q$ generates the correct BRST transformations on
$X ^m$, $c ^\alpha$ and $\bar c _\alpha$. Note that, since the
corresponding momenta vanish, we have eliminated $V ^\alpha$ and $B
_\alpha$ from the theory without affecting the Poisson brackets
between the remaining fields.

We note that our Hamiltonian (\ref{HAMI}) is the same as the one in
\cite{liraspsr},(modulo a total divergence) and consequently generate
the same equations of motion; those derivable from the action
(\ref{SGFS}). Furthermore, the BRST-charge (\ref{LBRS}) equals that in
\cite{liraspsr}, modulo equations of motion.

We may now quantize by making $X$, $c$ and $\Pi$ operators with
canonical commutation relations and introducing a Hermitean ordering,
all of which will be discussed in great detail below for the
light-cone gauge.  Finally, using the results of \cite{liraspsr}, we
have
\bbe{CHECK}
\left\{ {\hat Q,\hat Q} \right\}=0,
\eee
independent of the dimension $D$.

The straight-forward BRST-quantization descibed above is less
restrictive than requiring quantum conformal invariance in the
light-cone gauge. The consequences of the latter will be discussed in
great detail in the next section, but we want to point out that an
alternative would be to study the BRST quantization and implement the
requirement of conformal invariance there.  This could be done in
several ways.  In a covariant (as opposed to a light cone) formulation
one should check that there are BRST-invariant generators generating
the correct symmetry, and that $Q^2 = 0$. In a non-covariant
formulation the algebra proper has to be checked. Finally one might
study the BRST-quantization of a {\it conformally} covariant
formulation of the tensionless string.  This would essentially follow
the lines presented above, but with some additional constraints and
corresponding ghost/anti-ghost system.

\bigskip
\begin{flushleft}
{\bf 3.2 The Light-cone Operator Algebra}
\end{flushleft}
\bigskip

{\it Strategy:} Our aim is to determine whether it is possible to
realize the space-time conformal algebra in terms of quantum
operators, acting on a Hilbert space corresponding to physical
light-cone degrees of freedom only. Since the classical conformal
generators (\ref{x}) are polynomials of $X^i(\sigma)$ and
$P^i(\sigma)$, products of such fundamental operators are needed in
the quantum algebra, and we should regularize these products to make
them meaningful. In general the regularization involves some
arbitrariness, which hopefully can be parametrized by a finite number
of constants in the limit where the regulator is removed. For each
such constant there is a possible correction term to the composite
operator. The study of the relevant generalizations of the generators
can be organised by taking into account how they scale with the
regulator, $\varepsilon$, i.e. by power counting. We will demonstrate
how an anomaly in the space-time conformal algebra is unavoidable
under quite general and reasonable assumptions.

\bigskip
PRINCIPLES AND ASSUMPTIONS\\
\bigskip

{\it Locality:} Locality means that physical quantites at different
$\sigma$ contribute additively to the total observable, at least
classically implying that different space-time points also contribute
additively. If one ultimately wishes to construct an interacting
theory with some notion of space-time locality one should thus try to
respect world-sheet locality. It is therefore desirable to keep this
locality as manifest as possible in all arguments.  One may consider
two locality concepts: "covariant locality", which treats all
coordinates on an equal footing, or "transverse locality", which
treats only the transverse coordinates as fundamental.  The classical
conformal generators are integrals along the string of local
expressions depending on $X^i(\sigma)$, $P^i(\sigma)$ and
$X^-(\sigma)$. In fact, the integrands do not even contain
derivatives. They are clearly covariantly local.  However, as seen in
formula (\ref{gnutt},\ref{r}), $X^-(\sigma)$ depends on the whole
string if regarded as a function of $X^i(\sigma)$ and $P^i(\sigma)$,
and thus functions of $X^-(\sigma)$ are not in general transversely
local.

We shall assume that the quantized conformal generators are
covariantly local, since this is the locality concept satisfied
classically.  We write all expressions as functions of $X^i(\sigma)$,
$P^i(\sigma)$ and $X^- (\sigma)$, so that one can determine covariant
locality by inspection.  The interesting conformal generators are
cubic or quartic polynomials, so there is also a significant practical
advantage in keeping this form of the generators rather than
Fourier-transforming and thus obtaining multiple convolutions. The
interpretation of regularization will also be much clearer in our
approach. We require that any correction terms to the generators are
also covariantly local.

\bigskip
{\it Regularization:} In the classical case the physics can be studied
through functions on phase space, which in the light-cone gauge is
parametrized by canonical coordinates $X^i(\sigma)$, $P^i(\sigma)$ and
$x^-$, $p^+$. In quantum mechanics the classical canonical Poisson
brackets are simply replaced by commutators, but for non-linear
functions of the coordinates one also has to worry about ordering
problems, which for systems with an infinite number of degrees of
freedom may even involve divergencies. In our case the source of such
divergence problems is easily traced to the canonical commutation
relations
\bbe{CAN}
\lbrack x^-,p^+ ]=-i,\qquad \lbrack X^i(\sigma ),P^j(\sigma ') ]
=i\delta^{ij}\delta (\sigma-\sigma ')
\eee
and to our interest (for physical reasons) in local operators. For
example, the simple Hermitean operator $i X^i(\sigma) P^i(\sigma) - i
P^i(\sigma) X^i(\sigma)$ is directly seen to diverge.  Taking into
account that observables generally do not involve a precise value of
$\sigma$, but an integral over some region, this divergence may be
side-stepped by smearing $X^i(\sigma)$ and $P^i(\sigma)$, i.e. by
convoluting each of them with an approximate delta function. The
details of the approximate delta function will not matter, we only
assume
\bbe{af}
\mathop{\lim }\limits_{\varepsilon \to 0}\int {d\sigma f(\sigma )\delta
_\varepsilon
(\sigma )}=&f(0) \qquad
\delta _\varepsilon (-\sigma) &= \delta _\varepsilon (\sigma)\\
\int d\sigma \delta _\varepsilon (\sigma) =& 1 \qquad \quad
 \delta _{s\varepsilon} (\sigma) &= {1 \over s} \delta _\varepsilon
({\sigma \over s}), \eee where the regulator $\varepsilon$ is seen to
measure the scale of smearing. The limit $\varepsilon \to 0$ is the
local limit, which defines the physical system.

The regularization proposed above yields finite answers for all
commutators as long as $\varepsilon > 0$, but it is not the most
general kind of regularization even for monomial operators, since each
canonical coordinate factor is smeared independently of the others. A
general smearing function depending on the positions of all operators
correlate these positions, unless it is factorized into functions
depending only on one position each.  For such general regularizations
one may still measure the scale of smearing in terms of $\varepsilon$.

The quantity that defines the quantum algebra is the canonical
commutator, so regularizations of the special factorizable kind,
smearings of canonical coordinates, are easily handled. One just
calculates their effect on the canonical commutator, and proceeds from
there on, using a modified delta function in a regularized canonical
commutator:
\bbe{REGC}
\lbrack X^i(\sigma ),P^j(\sigma ') ]=
i\delta^{ij}\delta _\varepsilon (\sigma -\sigma ')
\eee
The smearing does not affect any other quantity. Obviously, a
physically sensible system can only be obtained in the local limit,
when the regulator is removed, but the presence of the regulator is
essential for making sense of intermediate steps in the calculations.
The more general kind of non-factorizable regularization is trickier,
but can be treated as giving rise to correction terms, small in the
parameter $\varepsilon$ that measures the regularization scale. This
is done by power counting. We shall in fact find that such corrections
cannot affect the conclusions we draw from simply regularizing the
canonical commutator. Before describing our power counting arguments
we list the other essentials of our procedure.

\bigskip
{\it Reference ordering:} Since the canonical commutators relate the
values of operator products with different orderings, there are
numerous ways of rewriting one and the same expression.  In an
algebraic calculation one wants to know whether further cancellations
are possible or not. By defining a standard reference ordering of
operators in all monomials, expressions can be directly compared, and
after cancellations the result is unique. In ordinary oscillator
calculations normal ordering is used, but since we do not have
oscillators, a different prescription has to be applied.

Defining a non-Hermitean version of the $M^{+-}(\sigma)$ current
\be
 {\cal M}(\sigma) \equiv p^+ X^-(\sigma)
\ee
and symbolically representing an operator and all its derivates raised
to an arbitrary power with the same letter, we have found the ordering
\bbe{refo}
p^+ X {\cal M} P
\eee
particularly convenient. It will be called the "reference ordering".
Note that $P^i(\sigma)$ annihilates the ground state, so that some
properties of normal ordering are retained in the tensionless limit
(cf. Appendix). The reference ordering is however used solely as a
book-keeping device, and our results will not be sensitive to the
choice of vacuum (unless one assumes a broken $\sigma \to -\sigma$
symmetry like in refs. \cite{gararual,ISBE}).

\bigskip
{\it Hermiticity:} Physical observables have real expectation values
and should be represented by Hermitean operators. Thus we should
require that the conformal generators (\ref{x}) are Hermitean. Hardly
any monomial operators ordered according to our reference ordering
(\ref{refo}) are Hermitean by themselves, so an ordered expression for
a generator contains correction terms that ensure Hermiticity.
Consider for example the operator $X^i(\sigma) P^i(\sigma) +
P^i(\sigma) X^i(\sigma)$ which is manifestly Hermitean, but not
ordered. Using the regularized canonical commutation relations
(\ref{REGC}) we find the ordered form $ 2 X^i(\sigma) P^i(\sigma) - i
C $, where the ordering constant $C$ diverges as $\varepsilon^{-1}$
when $\varepsilon \to 0$ due to the scaling behaviour (\ref{af}). Such
singular ordering terms are typical, but as will be shown below, there
will be {\it {only one independent ordering constant}} in the
conformal algebra, due to the algebraic relations between conformal
generators. For the power counting arguments we still have to keep
track of the $\varepsilon \to 0$ singularities from these ordering
terms.

\bigskip
{\it Conformal recursion:} The conformal algebra in light-cone
coordinates is
\bbe{L.C.I}
\lbrack p^i,k^j] &=&  m^{ij}-\delta^{ij}s\cr
\lbrack s,p^i] &=& p^i \qquad \qquad
\qquad\qquad \qquad \lbrack s,k^i] = -k^i\cr
\lbrack m^{ij},p^k] &=& \delta^{ik}p^j - \delta^{jk} p^i \qquad \ \qquad
\lbrack m^{ij},k^k] = \delta^{ik}k^j - \delta^{jk} k^i\cr
\lbrack m^{ij},m^{kl}]&=&
\delta^{ki}m^{jl}-\delta^{kj}m^{il}+\delta^{lj}m^{ik}-\delta^{li}m^{jk}
\nonumber
\eee
\bbe{L.C.II}
\lbrack s,p^\pm ] &=&  p^\pm \qquad \qquad \qquad \lbrack s,k^\pm ] =- k^\pm\cr
\lbrack k^i,p^\pm ] &=& -m^{\pm i} \qquad \ \qquad
\lbrack p^i,k^\pm ]
=m^{i\pm}\cr
\lbrack p^i,m^{j\pm}] &=& -\delta^{ij}p^\pm\cr
\lbrack m^{ij},m^{k\pm}]&=&
\delta^{ki}m^{j\pm}-\delta^{kj}m^{i\pm}\cr
\lbrack k^i,m^{j\pm}] &=&-\delta^{ij}k^\pm
\nonumber
\eee
\bbe{L.C.III}
\lbrack p^\pm ,m^{i\mp}] &=&-p^i\qquad \qquad \quad
\lbrack k^\pm ,m^{i\mp}]=-k^i\cr
\lbrack m^{i\pm},m^{j\mp}]&=&\delta^{ij}m^{\pm\mp}-m^{ij}\cr
\lbrack p^\pm,k^\mp]
&=&s\pm m^{+-}\nonumber
\eee
\bbe{L.C.}
\lbrack p^\pm,m^{+-}] &=&\mp p^\pm, \cr
\lbrack k^\pm ,m^{+-}] &=&
\mp k^\pm\cr
\lbrack m^{i\pm},m^{+-}]&=&
\mp m^{i\pm}
\eee
with {\it all other brackets vanishing}.  From this we can make
several interesting observations. \\ ({\it i}) The transverse
generators form a (D-2)-dimensional Euclidean conformal algebra by
themselves. The translation generators $p^i$ and the conformal boosts
$k^j$ are sufficient to span this algebra by repeated commutators.\\
({\it ii}) Adding $p^+$ and its repeated commutators with the
transverse generators produces $m^{i+}$, $m^{+i}$, and $k^+$.\\ ({\it
iii}) Finally adding $p^-$ then produces all remaining generators.\\

In following this three-step procedure of generating the full algebra,
one also encounters conditions relating the generators, so that an
erroneous ansatz for one of the initial generators $p^i, p^+, p^-$ or
$k^i$ will be revealed. If, on the other hand, all such conditions are
fulfilled, a realization of the conformal algebra has been
constructed.

In the present case we know the form of the classical generators, and
that their Poisson brackets satisfy the conformal algebra. The
quantized generators should be closely related to the classical, since
we expect there to be a classical limit when $\hbar \to 0$. However,
the quantum generators may deviate from the classical due to ordering
problems, due to the non-locality introduced via the regularization or
due to the renormalization of some quantities.

We may now take advantage of the structure of the algebra described
above and the form of the classical generators (\ref{x}). $p^i$, $p^+$
and $p^-$ are so simple that no ordering problems can affect them, and
$p^i$ and $p^+$ are linear so they are not even modified by
regularization. The ordered form of the Hermitian conformal boost
$k^i$ contains an ordering constant of order $\varepsilon^{-1}$. This
is the only independent ordering constant, since the commutator of two
Hermitean operators gives $i$ times a Hermitean operator, and all
other generators can be obtained in this way. It remains to study the
possible consequences of the regularization of $p^-$ and $k^i$.

Above we have discussed how ordering terms may appear in the
definition of the generators of the algebra. In addition, truly
serious ordering problems can arise because commutators of non-linear
operators need to be reordered before they comply with the fixed
reference ordering. This process may generate anomalous terms. By
constructing all generators in the recursive way described above we
disentangle the two problems: {\it Any deviation from the algebra
which cannot be absorbed in the redefinition of the generators is an
anomaly.}

\bigskip
{\it Conformal action on X:} Our object is to study whether the
conformal algebra works for the quantized tensionless string, and we
want to keep the geometrical picture of the transformations, not only
their algebra. To what degree can the conformal generators be modified
without jeopardizing their geometrical interpretation?  Classically
the conformal transformations are uniquely determined once a gauge has
been fixed completely, i.e. the action of the conformal group on the
coordinates of the string is precisely known when one has specified
these coordinates exactly. One such choice of gauge is the light-cone
gauge (though a rigid $\sigma$-translation is conveniently left
unfixed).  The philosophy of light-cone gauge quantization is to fix
the gauge classically and quantize the gauge fixed description of the
theory. This procedure should work whenever the geometry of the theory
is left untouched by quantization. In such a case the action of the
conformal generators is given by the classical expression except
possibly for correction terms vanishing as $\varepsilon \to 0$.

We assume that the geometry is left intact by quantization, and thus
that the action of the conformal group on $X^i(\sigma)$ approaches the
classical result when the regulator is removed.

\bigskip
{\it Power counting:} We assume that the quantum conformal generators
can be expanded in powers of the regulator $\varepsilon$ which has
dimension of world-sheet length. Higher powers of the regulator will
therefore combine with higher derivatives of the basic operators. As
an example of how this works consider a smeared version of $X^i(0)$:
\bbe{CONX}
X^i_\varepsilon (\sigma)&=&\int{d\sigma '{\delta_\varepsilon (\sigma -
\sigma ')X^i(\sigma ')}}\cr
&\approx &X^i(\sigma )+\textstyle{\varepsilon^2\over 2}{X^i}''(\sigma
)\int{ d\left({ {{\tilde \sigma}\over
\varepsilon}
}\right) \left({ {{\tilde \sigma}\over \varepsilon}}\right)^2
\delta_1\left({{{\tilde \sigma}\over \varepsilon}}\right)
}
\eee
There may also be singular terms in the conformal generators, but in
order to preserve the classical transformations of $X^i(\sigma)$, we
only allow such terms if they are required by Hermiticity and by the
reference ordering, as discussed above.

\bigskip
{\it Possible "counterterms":} We are now ready to state concretely
what correction terms to the conformal generators our principles
allow. It is sufficient to describe the possible corrections to $p^-$
and $k^i$, since they together with the trivial operators $p^+$ and
$p^i$ span the whole algebra.

In terms of reference ordered expressions the allowed correction terms
are:\\ ({\it i}) The singular term from ordering in $k^i$,
proportional to $ i\hbar \varepsilon^{-1} X^i$.\\ ({\it ii}) A
similar, but real and regular correction to $k^i$, proportional to
$\hbar X^i$.  Such a term also induces a c-number shift in the
dilatation operator $s$.\\ ({\it iii}) Various n'th-derivative terms
proportional to $\varepsilon^n$. The number of such possibilities is
constrained by dimensional analysis and by the requirement that they
should disappear in the classical limit. We refrain from classifying
them since we can instead show that their presence will not change our
conclusions.


\bigskip
CONFORMAL RECURSION RESULTS\\
\bigskip

Given an ansatz for the basic generators and a regularization we can
calculate the algebra, identify terms, and see what part of the result
survives in the local limit. Note that there are terms in the
generators diverging as $\varepsilon \to 0$. One has to consider also
terms vanishing in this limit, since they can combine with the
divergent terms.

The most important and most difficult part of the calculation depends
on the properties of $X^-(\sigma)$. This should not come as a
surprise, since the composite operator $X^-(\sigma)$ is obtained by
solving the non-abelian constraint (\ref{n},\ref{gnutt}). (It
generates the residual Virasoro symmetry (\ref{RTgen}).)  Thus, after
the gauge has been fixed, most of the non-trivial gauge symmetries of
a covariant formulation are encoded in this operator. In order to
complete the technical description of our method we therefore have to
describe how the properties of the operator $X^-(\sigma)$ are derived
and how it is regularized. Then we can finish the investigation by
deriving an anomalous term in the algebra spanned by uncorrected, but
quantized generators, and finally proving that no combination of
correction terms can conspire to cancel the anomalous contribution.

\bigskip{\it
$X^-$:} The constraint (\ref{n},\ref{gnutt}) which allows us to solve
for $X^-(\sigma)$ actually contains the derivative of $X^-(\sigma)$,
so we cannot get a closed expression for $X^-(\sigma)$ unless we
determine the integration constant.  However, for the conformal
algebra only the commutation rules of $X^- (\sigma)$ are needed, not
its precise form. The commutation rules can be found from integrating
the commutation relations of $X'^-(\sigma)$ calculable from
(\ref{mx}), and imposing the "sum-rule" that the zero-mode $x^-$
should be canonically conjugate to $-p^+$ and commute with the
transverse degrees of freedom.:
\bbe{SUM}
\int{d\sigma \left[{X^-(\sigma ),p^+}\right]}&=&-i\cr
\int{d\sigma \left[{X^-(\sigma ), X^i(\sigma ')}\right]} &=&
\int{d\sigma \left[{X^-(\sigma ), P_i(\sigma ')}\right]} =0
\eee
In addition one wants the commutation relations to be manifestly
periodic, since the physics is. Surprisingly, periodicity of the
commutation relations is not automatic, but it can be achieved by
making use of the one gauge symmetry that has been left unfixed in the
light-cone gauge, the rigid $\sigma$-translation $\Delta$. In the
following it will be convenient to phrase the discussion in terms of $
{\cal M}(\sigma) = p^+ X^-(\sigma)$. The remaining symmetry means that
\be
{\cal M}(1) - {\cal M}(0) = \int{d\sigma {\cal M}'(\sigma) } =
\int{d\sigma X'^i(\sigma ) P^i(\sigma )} \equiv \Delta = 0
\ee
in the physical phase space. In quantum mechanics the corresponding
statement is that $\Delta$ annihilates physical states, but this
"vanishing" of $\Delta$ is of course only effective if $\Delta$ has
been commuted to the right (or left) of any other operators. If one
reinterprets ${\cal M}'(\sigma)$ as
\be
{\cal M}'(\sigma) \to {\cal M}'(\sigma) - \Delta,
\ee
one automatically obtains
\be
 {\cal M}(1) - {\cal M}(0) = 0
\ee
as well as manifestly periodic commutation relations of ${\cal
M}(\sigma)$
\bbe{C-}
\lbrack X^i(\sigma ),X^-(\sigma ') ]&=&
{i\over p^+}X'^i(\sigma )h(\sigma -\sigma ')\cr
\lbrack P^i(\sigma ),X^-(\sigma ') ]&=&{i\over p^+}\left\{{P'^i(\sigma
)h(\sigma -\sigma ')+P^i(\sigma )\left[{1-\delta_\varepsilon (\sigma
-\sigma ')}\right]}\right\}\cr
\lbrack X^-(\sigma ),X^-(\sigma ') ]&=&{i\over p^+}\left\{{\left[{X'^-(\sigma
)+{\Delta\over p^+}}\right] h(\sigma -\sigma ')}\right.\cr
&-&\left.{\left[{X'^-(\sigma ' )+{\Delta\over p^+}}\right] h(\sigma '
-\sigma )}\right\}.
\eee
$h$ is given in eq. (\ref{ad}).  It is also necessary to understand
the effects of regularization on ${\cal M}(\sigma)$. Following the
principle of light-cone quantization we should keep the relation to
the transverse degrees of freedom as close as possible to the
classical relation.  Although it is not manifest, the regularized
${\cal M}'(\sigma)$ is in fact Hermitean, due to the $\sigma \to
-\sigma$ symmetry assumption in (\ref{af}). Other possible
regularization effects can be taken into account by allowing
correction terms in ${\cal M}'(\sigma)$, containing derivatives and
correspondingly being of higher order in $\varepsilon$. When
integrated to give ${\cal M}(\sigma)$ such terms could give
contributions to generators that do not necessarily appear to be
local. However, we should recall that we assumed covariant locality,
so that these correction terms should not appear isolated, but always
as parts of $X^-$. Therefore they can all be handled by determining
how they modify the commutation rules of ${\cal M}(\sigma)$, which
will include new terms with more derivatives than the classical terms.

\bigskip
{\it Anomaly:} Having fixed above what quantized generators are
physically acceptable, and how to commute them, we are now ready to
discuss the actual calculation of the quantum algebra.  The reference
ordering (\ref{refo}) is constructed so as to minimize the number of
times one has to reorder the result of a commutation in order to get
an ordered result. In many cases one can exclude any deviations from
the classical case, simply by checking that no new terms can appear
from reordering. In other cases, the procedure of {\it defining} the
more complicated conformal generators recursively from the algebra
absorbs ordering terms that would otherwise have appeared to be
dangerous. Only a few commutators of highly nonlinear generators
remain as possible sources of anomalies.

Reordering is necessary whenever on operator quadratic in position is
commuted with an operator quadratic in the conjugate momenta.  Due to
its construction as an integral of $X'^i(\sigma)P^i(\sigma)$ we may
regard ${\cal M}(\sigma)$ as linear in position and linear in
momentum. From the form of the generators (\ref{x}) and the algebra
(\ref{L.C.}) we then read off which commutators can cause problems.
One should note that the notorious Lorentz commutator
$[m^{i-},m^{j-}]$, which harbours the tensile string anomaly, not even
appears in the list of potential dangers. The reason is simply that
ordering problems for Hermitean operators arise at higher order in
$\hbar$ when the ordering is defined in terms of positions and
momenta, than when it is defined in terms of annihilation and creation
operators, a fact that is familiar from the one-dimensional harmonic
oscillator.  Postponing for a while the discussion of commutators with
the quartic generator $k^-$ we are left with only $[m^{i-},k^+]$ and
$[m^{i-},k^j]$.

The first of the above commutators can be checked to give the correct
answer, while the second splits naturally in two parts.  From the
conformal algebra (\ref{L.C.}) we see that the trace part defines
$k^-$, but a trace-less contribution would necessarily be anomalous.
Indeed, this is what happens. A lengthy commutator calculation,
involving also replacements $X'^i(\sigma)P^i(\sigma) \to {\cal
M}'(\sigma)$ and integrations by parts, finally yields an anomaly:
\bbe{ANO}
\left[{
k^i,m^{j-} }\right] +\delta^{ij}k^-&\propto&
\int{
d\sigma
\left({
{1\over{\varepsilon p^+}} }\right)
\left\{{
X^i(\sigma )P^j(\sigma )-x^ip^j+h.c.  }\right\}_{t.l.} }\cr &\equiv&
\left({
{1\over{\varepsilon p^+}} }\right)L^{ij}_{t.l.},
\eee
where the subscript {\it t.l.} denotes the trace-less part.  We note
that the anomaly (\ref{ANO}) vanishes for $\sigma$-independent
coordinates and momenta. This feature is shared by all the other
anomalies, generated by commutators with $k^-$, and therefore all
quantization problems are caused by the extendedness of the string.
In contrast, we find that massless scalar {\it point-particles},
described by the constant zero-modes, {\it do} admit a quantum
conformal symmetry.

The anomalous term is divergent, a state of affairs which is not
familiar from relativistic field theory, but the present framework is
quite different. One might still ask whether there is any possibility
of removing the anomaly by modifying the short-distance behaviour of
the theory. In the next section we will argue that there is no way of
doing this while respecting the natural locality assumptions we have
discussed at length above.

\bigskip
{\it Anomaly cancellation?:} There are basically two ways of modifying
the generators when taking quantum short-distance effects into
account. Either generators are simply rescaled (renormalized) due to
short-distance effects, or the generators receive corrections from
derivative terms.

Rescaling the generators is quite unnatural, since there is no
dimensionless coupling in the present problem, and thus no limit where
the rescaling could be small. There is also an algebraic reason why
this alternative is ruled out. One may check that the only rescalings
of the generators that preserve the conformal algebra and the
definition of the Hamiltonian (\ref{LCH}) are those that are generated
inside the algebra by $m^{+-}$. Hence the anomaly does not have a form
which allows it to be scaled away.

One could also envisage that the anomaly could be absorbed by changing
some string model parameters. At our disposal we have, at most, a
string tension $T$ and the speed of light $c$, but changing these
parameters deform the algebra too drastically, already at the
classical level.

It remains to consider derivative corrections, together with
appropriate powers of $\varepsilon$. Since also negative powers of
$\varepsilon$ appear in the generators, such terms can in principle
correct the algebra in the local limit. However, among the commutation
rules for the basic operators (\ref{CAN},\ref{C-}) there are none that
decrease the number of derivatives, and thus the anomaly (\ref{ANO})
can never be compensated by local modifications of generators.

\bigskip
{\it Open strings:} We have checked that an analogous structure of
anomalies appears for the open string with boundary conditions
$V^1(0)=V^1(1)=0$. The commutators of $X'^-(\sigma)$ are modified
since they should no longer be periodic, but the end result and the
arguments excluding compensating terms are unchanged.

\bigskip
DISCUSSION\\
\bigskip

Faced with the anomaly (\ref{ANO}) we have a number of alternatives.

({\it i}) We simply accept that space-time conformal symmetry is not a
symmetry of the quantum theory. Global symmetries can be anomalous
without ruining the consistency of the theory, and the Poincar\'e
subgroup is still a symmetry.  But it is then quite mysterious why the
space-time symmetry group of the tensile string should happen to give
the right criterion for the critical dimension{\footnote{Recall that
there is a formulation (\cite{akul,ISBE}) of the tensionless string
where conformal symmetry is manifest before fixing the gauge, just as
the Lorentz symmetry before fixing the light-cone gauge in the tensile
string.}. The fact that the conformal symmetry seems to be present in
lieu of the 2-D Weyl symmetry of the tensile string also suggests that
it should be taken more seriously.

({\it ii}) Something has been overlooked in going from the geometrical
action to the quantized gauge-fixed theory. Perhaps purely geometrical
and auxiliary fields as $V^{\alpha}$ get a life of their own, like the
conformal factor of the metric for non-critical tensile strings. Such
a state of affairs is quite possible, but it is also hard to reconcile
with the picture of tensionless strings being the $T \to 0$ limit of
the ordinary string. How could new degrees of freedom suddenly appear?
In any case, we cannot discuss this alternative in more detail with
the tools of the present article.

({\it iii}) The anomalies actually vanish. This may happen because
their operator {\it form} is not the whole story. One also has to take
the Hilbert space they act on into account. Previously we have only
demanded that the states should be invariant under the rigid
$\sigma$-translation gauge symmetry, and in the appendix we discuss
what conditions the vacuum should satisfy in quite general terms.  We
also know that a Fock space is not appropriate, since oscillators and
decomposition into positive and negative frequencies does not describe
free particles, which is roughly what the tensionless string consists
of.

We should thus ask whether we can regard the vanishing of the
anomalies as conditions on a Hilbert space yet to be constructed. In a
relativistic field theory we would know the free Hilbert space and
could thus not take such an attitude, but here we are faced with a new
situation and should at least check the implications of such an
unconventional treatment. There is however reason to fear that
regarding the anomalies as constraints on the Hilbert space reduces
its "size" considerably. We shall return to this question in the
conclusions, but we first describe the consequences of our
assumptions.

The anomaly (\ref{ANO}) is proportional to the generators of
$\sigma$-dependent {\it special linear transformations} of the
transverse coordinates ($L^{ij}_{t.l.}$ is trace-less). The constraint
\bbe{AINT}
\tilde L^{ij}_{t.l.}|{PHYS}>=0,
\eee
is imposed to restore quantum conformal symmetry, so one should also
require its variation under conformal transformations to annihilate
physical states. The full set of constraints is given by the anomaly
(\ref{ANO}) and the additional anomalies from $[k^+,k^-]$, $[k^i,k^-]$
and $[m^{i-},k^-]$, together with their repeated commutators with each
other and with the conformal generators. We have checked part of this
algebra using Mathematica, and found that one obtains all the
generators of $\sigma$-dependent special linear transformations
\bbe{ZM}
X^m(\sigma )P^n(\sigma)-x^mp^n +h.c.
\eee
not just the transverse components. Some automatically vanish in the
light-cone gauge, but the fact that all non-zero generators appear
ensures that the constraints have a covariant interpretation. (This
would have been evident already from the construction, were it not for
the special role of zero-modes.)

Phrased covariantly, we could thus state that the Hilbert space of
tensionless string states should be invariant with respect to special
linear transformations on the canonical coordinates, zero-modes modes
excluded, {\it and} invariant with respect to all conformal
transformations of these linear transformations. The full algebra of
constraints is then generated by repeated commutators.

At this point it is illuminating to recall a result due to Ogievetsky
\cite{og}: The finite set of special linear transformations and conformal
transformations together generate all (analytic) diffeomorphisms.
Along the lines suggested by this result one can speculate that
physical states annihilated by the constraints have a non-zero mode
dependence that is invariant under analytic space-time
diffeomorphisms. They should then be characterized by the topology of
the loop they trace in space, e.g. by the number of
self-intersections, and possibly also by discontinuites in
derivatives, kinks and cusps.

The algebra of constraints could give this result, and it does give
something very similar, but the separation of zero-modes is quite
subtle, and it may turn out that some additional non-topological
structure survives. The problem is that the conformal transformations
couple zero-modes and $\sigma$-dependent modes. Most of this coupling
seems to disappear when all constraints are taken into account
simultaneously, but we have not yet found a conclusive answer to how
the zero-modes affect the precise symmetry of the Hilbert space.

While the above considerations could enlarge the Hilbert space
compared to a purely topological theory, some of the anomalies have a
form that seem to reduce it even more. In particular, $[k^+,k^-] = 0$
gives rise to a constraint
\bbe{XX}
\int{
d\sigma (X^i(\sigma) - x^i) (X^i(\sigma) - x^i)|{PHYS}> = 0 }
\eee
which in its turn, through commutators with $p^-$, generates
\bbe{XP}
\int{
d\sigma (X^i(\sigma) - x^i) (P^i(\sigma) - p^i)|{PHYS}> = 0 }\cr
\int{
d\sigma (P^i(\sigma) - p^i) (P^i(\sigma) - p^i)|{PHYS}> = 0} .
\eee
Since two of the above equations look like sums of squares, it is
tempting to draw the conclusion that $X^i(\sigma) - x^i = P^i(\sigma)
- p^i = 0$, i.e.  that the physical Hilbert space should only consist
of zero-modes.  However, since the operators are regulated, the
factorization tacitly assumed in this argument is not strictly true.
In any case, the state space is severely restricted. In fact, the
constraints (\ref{XX},\ref{XP}) illustrate how one reobtains some of
the constraints lost through the zero-mode subtleties. Namely, by
supplying them with appropriate powers of $p^+$, they are identical
with the contributions from the $\sigma$-dependent modes to $k^+$, $s$
and $p^-$.
\bigskip

\begin{flushleft}
\section{Conclusions}
\end{flushleft}
\bigskip

      We have discussed classical and quantum aspects of the
tensionless string. At the classical level we have reported on a
particular method for deriving tensionless strings (and tensionless
$p$-branes) which leads to an action with auxiliary fields that have a
geometric meaning, just as for the tensile Weyl-invariant string.

In describing the symmetries of the model we have placed special
emphasis on the space-time conformal symmetry, a symmetry that we
think of as replacing Weyl-invariance in the $T\to 0$ limit. In this
context it is interesting to note that in quantum field theory it is
often useful to understand Poincar\'e \ symmetry as a broken conformal
symmetry. In the early days of supergravity, e.g.,the appearence of
the minimal set of auxiliary fields $N=1$ Poincar\'e \ supergravity
became clear after when that theory was viewed as broken conformal
supergravity. From this point of view it is thus natural to try to
gain insight into the Poincar\'e \ invariant tensile string theory by
comparing it to a tensionless conformally invariant string theory.
This complements the motivation from high-energy string theory
presented in the introduction.

Quantizing the theory we find that the quantum theory differs from the
classical theory drastically. Either the symmetries are changed or the
degrees of freedom are changed. This conclusion should be independent
of our method, and it is certainly independent of the choice we then
make: For reasons described in the discussion section we have pursued
the idea that the classical symmetries should survive as quantum
symmetries. A careful analysis leads us to conclude that this is only
possible if the physical Hilbert space is drastically constrained. We
seem to be left with only diffeomorphism singlets as physical states.
This points to a topological theory in quite an unsuspected way.

Note however that we have {\it not} constructed the Hilbert space,
only derived restrictions on it.

Some of the constraints (\ref{AINT},\ref{XP}) indicate that the
physical states should be massless, spinless and have zero scaling
dimension. These constraints remove the model very far from what is
expected from the classical theory, but there is circumstantial
evidence that we are still on the right track.

Our result is consistent with the selection rules found by Gross
\cite{GROS1} for high-energy tensile string scattering. He found that
scattering amplitudes for all external states where directly related
to the tachyon amplitudes by kinematic factors in the $T \to 0$ limit.
Thus a description in terms of very few states may capture the
essential physics. In addition these amplitudes are given by
polarizations (spin) in the scattering plane, i.e. the plane defined
by the relative momenta. Other polarization directions do not affect
the amplitudes.  The constraints (\ref{AINT}) imply that no spin is
allowed for a single tensionless string, but on the other hand spin
does not seem to make a difference unless it is combined with a
relative momentum. The ultimate test of the constraints must come from
how they affect multi-string Hilbert spaces in an interacting theory
of tensionless strings.

Another sign of a drastic reduction of the spectrum is the scaling
argument of Atick and Witten \cite{ATIC1}, which indicates that the
short-distance degrees of freedom of string theory are much fewer than
in particle theory. Finally, we can get an idea of a physical origin
of the constraints by comparing with the study of Karliner et al.
\cite{KARL1}, on the wavefunction of the ordinary string. At distances
below a fundamental length $T^{-1/2}$, fluctuations completely
dominate the wavefunction, and it makes little sense to specify a
particular string configuration. In the tensionless limit this
behaviour should extend to all of space-time.

\bigskip
\begin{flushleft}
{\bf Acknowledgements:} We thank M.B. Green, J. Grundberg, H. Hansson,
C. Hull, S. Hwang, G. Papadopoulos and M. Ro\v cek for useful comments
and fruitful discussions.
\bigskip
\end{flushleft}

\eject
\begin{flushleft}
\section{Appendix}
\end{flushleft}

\bigskip
{\bf The Vacuum}
\bigskip

We define the vacuum $\left| { 0 }\right\rangle$ for the $T\to 0$
theory by the condition that
\bbe{deva}
P^m\left| { 0 }\right\rangle _0&=&0.
\eee
$X^m(\sigma)$ and $P^m(\sigma)$ are the position and momentum
operators at $\tau =0$ used to quantize the theory. In terms of their
Fourier components the condition (\ref{deva}) reads
\be
P^m_{\bf n}\left| { 0 }\right\rangle _0&=&0
\ee
while
\be
X^m _{\bf n}\left| { 0 }\right\rangle _0&\not=&0.
\ee
At $\tau =0$ the position and momentum operators can be identified
with those of the tensile theory, although the models have different
dynamics. We have arrived at the definition (\ref{deva}) guided by a
wish to keep a relation to the $T\not =0$ Hilbert space and vacuum. We
have argued as follows:

The vacuum $ \left| { 0 }\right\rangle _T $ for the tensile theory is
annihilated by the oscillators
\bbe{TVAC}
\alpha^m _{\bf n}(T) \left| { 0 }\right\rangle _T
=\tilde \alpha^m _{\bf n}(T) \left| { 0 }\right\rangle _T=0
\qquad\forall {\bf n}>0.
\eee
The oscillators may be expressed in terms of of the Fourier components
of the coordinate and momentum operators as follows:
\bbe{osci}
\alpha^m _{\bf n}(T) &=&-in\sqrt TX^m _{\bf n} +{1 \over {2\sqrt T}}P^m _{\bf
n}\cr
\tilde\alpha^m _{-\bf n}(T) &=&in\sqrt TX^m _{\bf n} +{1 \over {2\sqrt T}}P^m
_{\bf n}.
\eee
We want to maintain a connection to $\left| { 0 }\right\rangle _T$
when we define the vacuum $\left| { 0 }\right\rangle _0$ for the
tensionless theory. As a first attempt we try defining $\left| { 0
}\right\rangle _0$ in analogy to (\ref{TVAC}), i.e.,
\bbe{def1}
\alpha^m _{\bf n}(T) \left| { 0 }\right\rangle _0&=&0\cr
\tilde \alpha^m _{\bf m}(T) \left| { 0 }\right\rangle _0&=&0
\eee
with perhaps a different range of ${\bf m}$ and ${\bf n}$. If it is to
hold in the limit $T\to 0$, (\ref{osci}) implies
\bbe{Pres}
P^m_{\bf n} \left| { 0 }\right\rangle _0 =P^m _{-\bf m} \left| { 0
}\right\rangle _0=0
\eee
since $X^m$, $P_m$ and $\left| { 0 }\right\rangle _0 $ are
$T$-independent by assumption.  But then by (\ref{def1}) we also have
\be
X^m_{\bf n} \left| { 0 }\right\rangle _0 =X^m _{-\bf m} \left| { 0
}\right\rangle _0=0.
\ee

If we assume (\ref{def1}) to hold for all ${\bf m},{\bf n} >0$, we
thus have $X^m_{\bf n} \left| { 0 }\right\rangle _0=P^m_{\bf n} \left|
{ 0 }\right\rangle _0=0$ for all ${\bf n}\not =0$. This is
inconsistent with the commutation relations.

If we assume instead that (\ref{def1}) holds for all ${\bf n}>0$ and
for all ${\bf m}<0$, we have $X^m_{\bf n} \left| { 0 }\right\rangle
_0=P^m_{\bf n} \left| { 0 }\right\rangle _0=0$ for all ${\bf n}>0$.
This is the choice made in \cite{gararual}. One expects, however, that
an asymmetric treatment of $\alpha$ and $\tilde \alpha$ should lead to
a breakdown of the global $\sigma$-translational symmetry and hence to
a non-zero two-momentum $P_\sigma$ for the closed tensionless string.
This is indeed what is reported in \cite{gararual}.

We further regard the possibility of "mixed choices", i.e., some
positive and some negative $\bf n$'s and $\bf m$'s as completely
unnatural.

Having thus discussed and discarded the possibility of requiring
(\ref{def1}) to hold for $T\not =0$ , we turn to the remaining option;
that this is satisfied in the limit $T\to 0$ only. We still find the
restriction (\ref{Pres}), of course, but $X^m _{{\bf m},{\bf n}}\left|
{ 0 }\right\rangle $ is not determined. This leaves us with
(\ref{deva}).

\eject


\begin{thebibliography}{99}

\bibitem{AMAT3}
{D. Amati,} {M. Ciafaloni,} and G. Veneziano, {\it Phys. Lett.} {\bf
197B} (1987) 81; {\it Int. J. Mod. Phys.} {\bf A3} (1988) 1615; {\it
Phys. Lett.} {\bf 216B} (1989) 41.

\bibitem{MUZI1}
I.J. Muzinich and M. Soldate, {\it Phys. Rev.}, {\bf D37} (1988) 359.

\bibitem{BOSU2}
B. Sundborg, {\it Nucl. Phys.} {\bf B306} (1988) 545.

\bibitem{GROS2}
D.J. Gross and P.E. Mende,
\newblock {\it Phys. Lett.} {\bf 197B} (1987) 129; {\it Nucl. Phys.}
{\bf B303} (1988) 407.

\bibitem{GROS1}
D.J. Gross,
\newblock {\it Phys. Rev. Lett.} {\bf 60} (1988) 1229.

\bibitem{GROS4}
D.J. Gross and J. Manes,
\newblock {\it Nucl. Phys.} {\bf B326} (1989) 73.

\bibitem{ALVA1}
E. Alvarez,
\newblock {\it Phys. Rev.} {\bf D31} (1985) 418.

\bibitem{BOSU1}
B. Sundborg,
\newblock {\it Nucl. Phys.} {\bf B254} (1985) 583.

\bibitem{BOWI1}
M.J. Bowick and L.C.R. Wijewardhana,
\newblock {\it Phys. Rev. Lett.} {\bf 54} (1985) 2485.

\bibitem{ATIC1}
J.J. Atick and E. Witten,
\newblock {\it Nucl. Phys.} {\bf B310} (1988) 291.

\bibitem{WITT1}
E. Witten,
\newblock {\it Comm. Math. Phys.} {\bf 117} (1988) 353.

\bibitem{WITT2}
E. Witten,
\newblock {\it Phys. Lett.} {\bf 206B} (1988) 601.

\bibitem{WITT3}
E. Witten,
\newblock {\it  Comm. Math. Phys.} {\bf 118} (1988) 411.

\bibitem{LABA1}
{J. Labastida,} {M. Pernici,} and E. Witten,
\newblock {\it  Nucl. Phys.} {\bf B310} (1988) 611.

\bibitem{WITT4}
E. Witten
\newblock {\it  Phys. Rev. Lett.} {\bf 61} (1988) 670;
{\it Phil.Trans.R.Soc.London.A} {\bf 329} (1989) 349.

\bibitem{BIRM1}
{D. Birmingham,} {M. Blau,} {M. Rakowski} and G. Thompson,
\newblock {\it Phys. Rep.} {\bf 209} (1991) 129

\bibitem{jiulbs}
{J. Isberg,} {U. Lindstr\"om} and B. Sundborg,
\newblock {\it Phys. Lett.} {\bf 293B} (1992) 321.

\bibitem{KARL1}
{M. Karliner,} {I. Klebanov,} and L. Susskind,
\newblock {\it Int. J. Mod. Phys.} {\bf 3A} (1988) 1981.

\bibitem{akul}
A. Karlhede and U. Lindstr\"om, {\it Class.Quant.Grav.} {\bf 3} (1986)
L73.

\bibitem{ulbsgt1}
U. Lindstr\"om, B. Sundborg and G. Theodoridis, {\it Phys.Lett.} {\bf
253B} (1991) 319.

\bibitem{ulbsgt2}
U. Lindstr\"om, B. Sundborg and G. Theodoridis, {\it Phys.Lett.} {\bf
258B} (1991) 331.

\bibitem{roli}
U. Lindstr\"om and M. Ro\v cek, {\it Phys.Lett.} {\bf 271B} (1991) 79.

\bibitem{zh}
A.A. Zheltukhin, {\it Sov.J.Nucl.Phys.} {\bf 48} (1988) 375.

\bibitem{banerual}
A. Barcelos-Neto and M. Ruiz-Altaba, {\it Phys.Lett.} {\bf 228B}
(1989) 193.

\bibitem{liraspsr}
F. Lizzi, B. Rai, G. Sparano and A. Srivastava, {\it Phys.Lett.} {\bf
182B} (1986) 326.

\bibitem{gararual}
J. Gamboa, C. Ramirez and M. Ruiz-Altaba, {\it Nucl.Phys.} {\bf B 338}
(1990) 143.

\bibitem{sc}
A. Schild, {\it Phys.Rev.D} {\bf 16} (1977) 1722.

\bibitem{Gross-Per}
D.J. Gross and V. Periwal, {\it Phys.Rev.Lett.} {\bf 60} (1988) 2105.

\bibitem{mass-shift}
B. Sundborg, {\it Nucl.Phys.} {\bf B319} (1989) 415.

\bibitem{HASS} S. Hassani, U. Lindstr\"om and R. v.Unge, {\it "Classically
Equivalent Actions for Tensionless p-branes"}, University of Stockholm
preprint USITP-93-14 (1993).

\bibitem{BRIN} L. Brink, P. diVecchia and P.S. Howe, {\it Phys. Lett.}
{\bf B65} (1976) 435.

\bibitem{DESE} S. Deser and B. Zumino, {\it Phys. Lett.} {\bf B65}
(1976) 369.

\bibitem{ggrt}
P. Goddard, J. Goldstone, C. Rebbi and C.B. Thorn, {\it Nucl.Phys.}
{\bf B56} (1973) 109.

\bibitem{og}
V.I. Ogievetsky, {\it Lett.Nuovo Cim.} {\bf 8} (1973) 988.

\bibitem{ISBE}
J. Isberg, {\it "Tensionless Strings with Manifest Space-Time
Conformal Invariance"} {USITP-92-10} (1992).

\bibitem{HWAN}
S. Hwang, {\it Phys.Rev.} {\bf D28} (1983) 2614.

\bibitem{KATO}
M. Kato and K. Ogawa, {\it Nucl.Phys.} {\bf B212} (1983) 443.


\end{thebibliography}
 \end{document}